\documentclass[a4paper,11pt]{article}

\usepackage{jinstpub} 
\usepackage{lineno}

\title{\boldmath Supernova Neutrino Detection in LZ}

\author[a,1]{D.~Khaitan,\note{Corresponding author.}}
\affiliation[a]{University of Rochester, Department of Physics and Astronomy, Rochester, NY 14627, USA}
\emailAdd{dkhaitan@u.rochester.edu}

\abstract{In the first 10 seconds of a core-collapse supernova, almost all of its progenitor's gravitational potential, O(10$^{53}$~ergs), is carried away in the form of neutrinos. These neutrinos, with O(10~MeV) kinetic energy, can interact via coherent elastic neutrino-nucleus scattering (CE$\nu$NS) depositing O(1~keV) in detectors. In this work, we demonstrate that low-background dark matter detectors, such as LUX-ZEPLIN (LZ), optimized for detecting low-energy depositions, are capable of detecting these neutrino interactions. For instance, a 27~M$_\odot$ supernova at 10~kpc is expected to produce $\sim$350 neutrino interactions in the 7-tonne liquid xenon active volume of LZ. Based on the LS220 EoS neutrino flux model for a SN, the Noble Element Simulation Technique (NEST), and predicted CE$\nu$NS cross-sections for xenon, to study energy deposition and detection of SN neutrinos in LZ. We simulate the response of the LZ data acquisition system (DAQ) and demonstrate its capability and limitations in handling this interaction rate. We present an overview of the LZ detector, focusing on the benefits of liquid xenon for supernova neutrino detection. We discuss energy deposition and detector response simulations and their results. We present an analysis technique to reconstruct the total number of neutrinos and the time of the supernova core bounce.}


\collaboration[c]{on behalf of the LZ Collaboration}

\proceeding{LIDINE 2017: Light Detection in Noble Elements\\
22-24 September 2017\\
SLAC National Accelerator Laboratory}

\begin{document}
\maketitle
\flushbottom

\section{Introduction}
\label{sec:intro}

During a core-collapse supernova (SN), $\sim$99~\% of the progenitor's gravitational potential energy is converted into neutrino flux emitted within 10~seconds, making supernovae among the most energetic events in the universe~\cite{ref1,ref2,ref3}. Despite recent progress we are still far from understanding the underlying physical processes of a SN and the role neutrinos play during them~\cite{ref4,ref5,ref6}. A high-statistics detection of all neutrino flavors will be essential to reconstruct global emission properties, such as the total explosion energy emitted~\cite{Drukier,Beacom,Horowitz}.

Coherent elastic neutrino nucleus scattering (CE$\nu$NS) was detected for the first time by the COHERENT Collaboration at the Spallation Neutron Source~\cite{Coherent}. This interaction, mediated by the Z-boson, is equally sensitive to all flavors of neutrinos making it a promising channel to detect neutrinos from a SN. The nature of the interaction carries numerous implications for experiments that intend to detect SN neutrinos via CE$\nu$NS: the neutrino emission curve could be reconstructed without uncertainties that arise from neutrino oscillation~\cite{SnNuFutureChakraborty}; and, these detectors would be able to reconstruct the total neutrino emission and will complement experiments that are only sensitive to $\overline{\nu_e}$ and $\nu_e$~\cite{TimeRecon}.

Tens-of-milliseconds after the collapse of the inner core is halted, the implosion rebounds and bounces outward and a bright neutronization or breakout burst dominated by $\nu_e$  from electron capture occurs. During neutronization, the average energy of the emitted neutrinos is between 9~-~12~MeV. This burst is followed by an accretion phase, tens to hundreds of milliseconds long, over which $\overline{\nu_e}$ and $\nu_e$ dominate with the average neutrino energy rising to $\sim$15~MeV. The next phase is cooling, which lasts a few tens of seconds, during which the core sheds most of its gravitational binding energy and the average neutrino energy steadily decreases to $\sim$6~MeV. During this phase, neutrino-antineutrino pairs dominate neutrino production, and luminosities and temperatures gradually decrease. The neutrino flux is equally divided among flavors during cooling and carries away half the total gravitational energy of the progenitor ~\cite{ref3, SuNuSignal1, SuNuSignal2, SuNuSignal3, SuNuSignal4, KateScholbergNu}. The spectral energy distribution of the emitted neutrinos varies from phase to phase of the SN with the distribution most sharply peaked during neutronization and broadest during cooling~\cite{ref3,NuSpectFit}. To neutrinos in this energy range, the nucleus appears as a coherent target with a cross-section enhanced by the square of its nucleon number~\cite{ref3,ref6,Drukier,Beacom,Horowitz}. However, there is a trade-off between nucleus mass and energy transfer; as the nucleus mass increases the energy transfer decreases. Thus, to detect neutrinos emitted from SN a low-threshold detector with heavy nuclei is desired. The LUX-ZEPLIN (LZ) experiment is a dark matter experiment that will be sensitive to neutrino interactions from SN events.

In Section~\ref{sec:lz}, we present an overview of the response of the LZ detector to neutrinos emitted from a SN. We use neutrino emission curves from Ref.~\cite{ref3,LS220EOS} for a 27 M$_\odot$ progenitor and simulate the response of the data acquisition system (DAQ). In section~\ref{sec:analysis}, we present the performance of the LZ DAQ to these high interaction rate events by varying the distance to the SN. We then demonstrate a fitting method that is applied to the recorded trigger time information to reconstruct the total neutrino interaction rate and the time of the bounce. 

\section{SN Neutrino Signal in LZ}
\label{sec:lz}

\begin{figure}[htbp]
\centering 
\includegraphics[width=.45\textwidth]{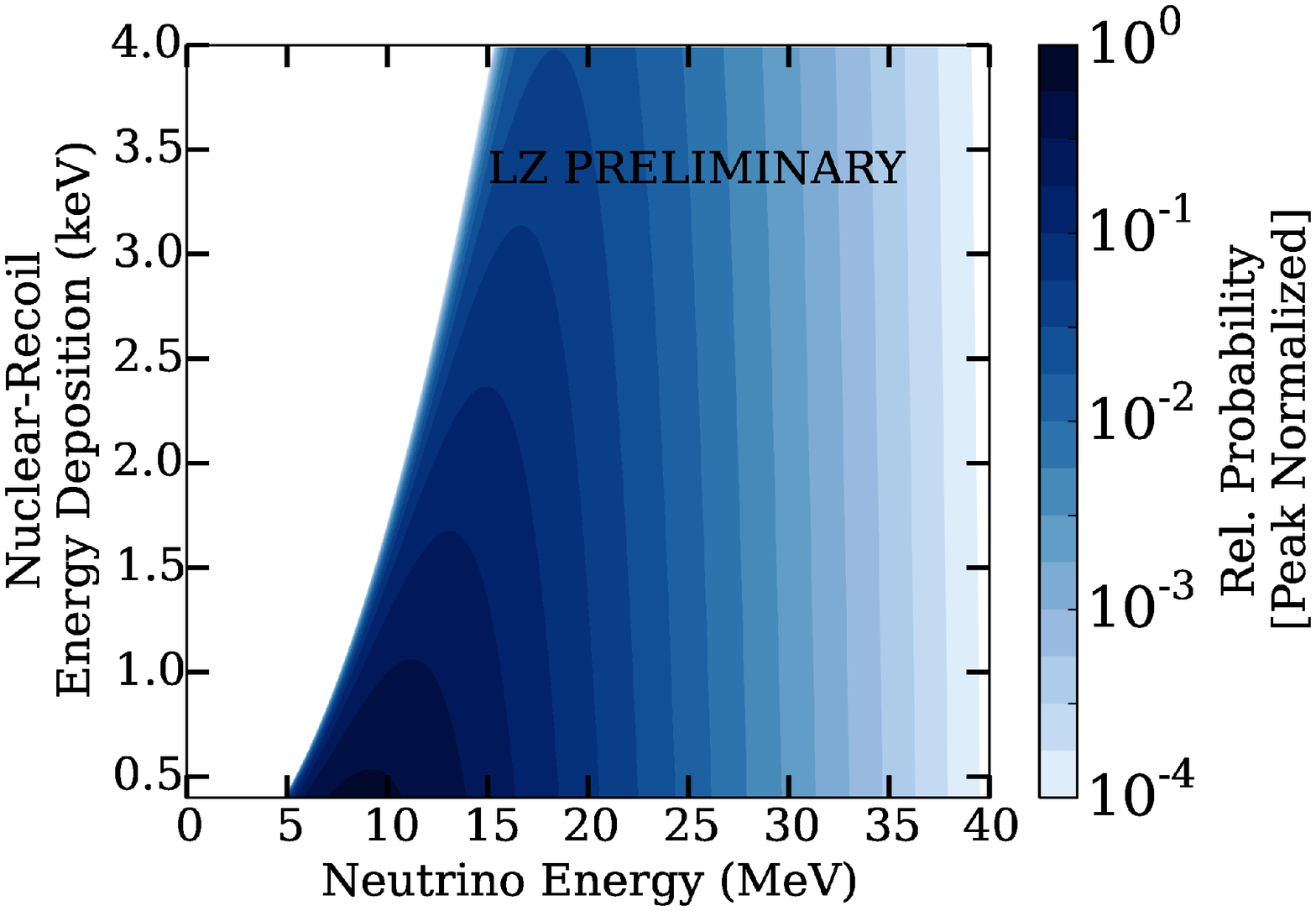}
\qquad
\includegraphics[width=.45\textwidth]{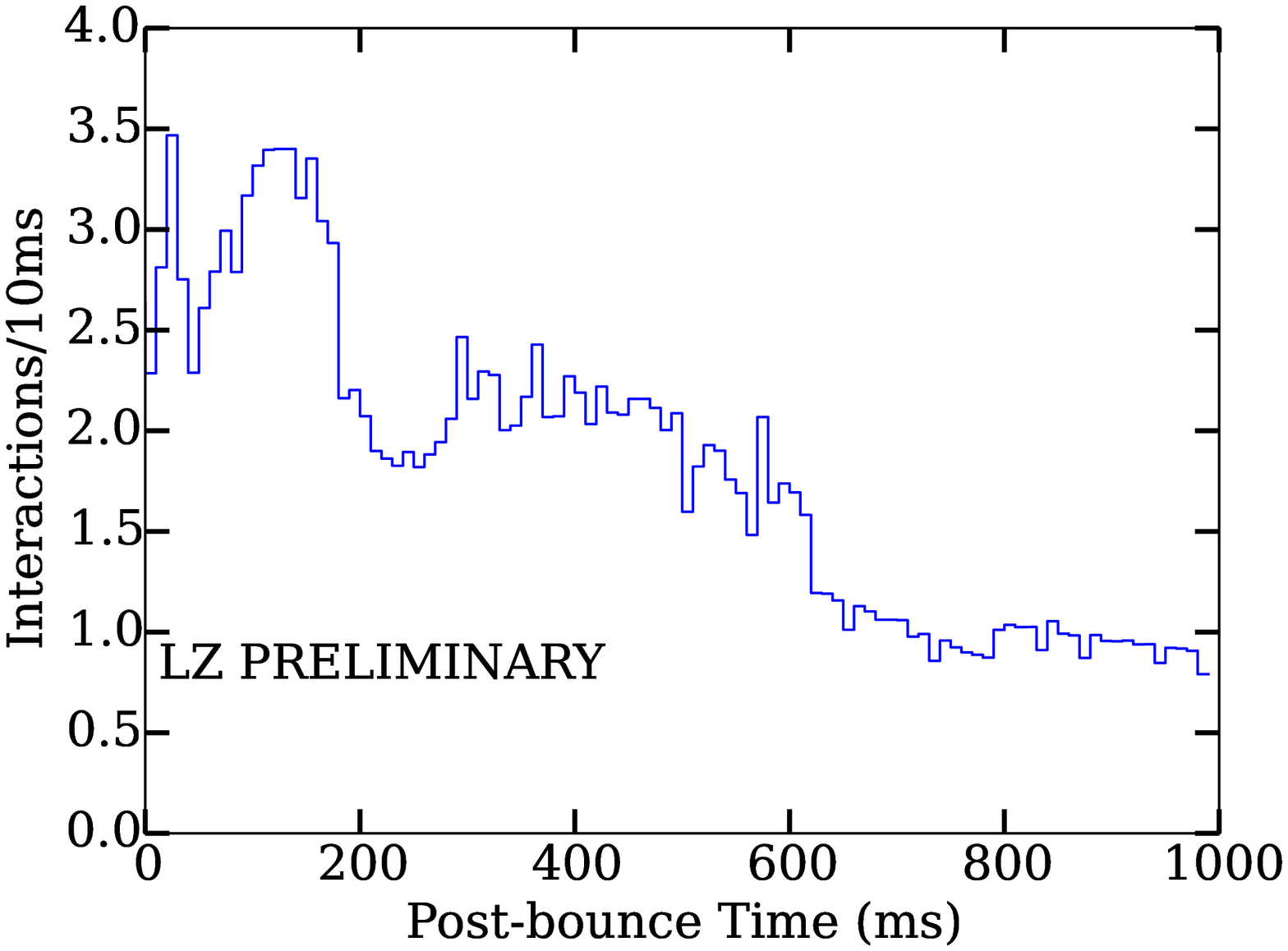}
\caption{\emph{(Left)} The relative probability distribution for a neutrino energy deposition in liquid xenon. We present the case where the incident neutrino flux has an average energy of 10~MeV and a spectral profile from the accretion phase~\cite{ref3, Drukier, Horowitz,NuSpectFit}. \emph{(Right)} The expected neutrino interaction rate in LZ from a 27~M$_\odot$ SN at 10~kpc assuming a detection energy threshold of 0.5 keV. About 184~$\pm$~13 neutrino interactions are expected in the first second and 357~$\pm$~19 in total. }
\label{fig:energy_dep}
\end{figure}

\begin{figure}[htbp]
\centering 
\includegraphics[width=.45\textwidth]{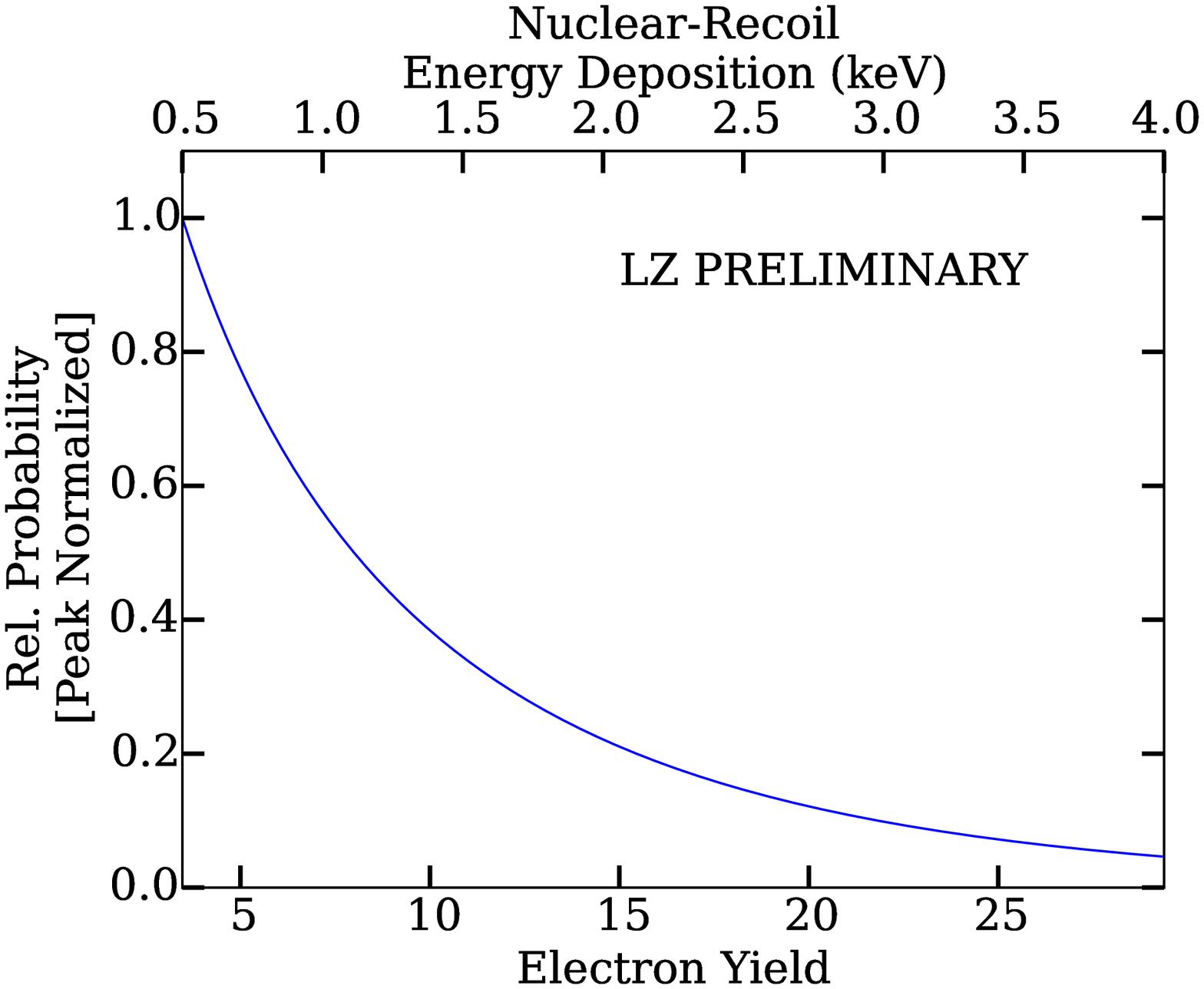}
\qquad
\includegraphics[width=.45\textwidth]{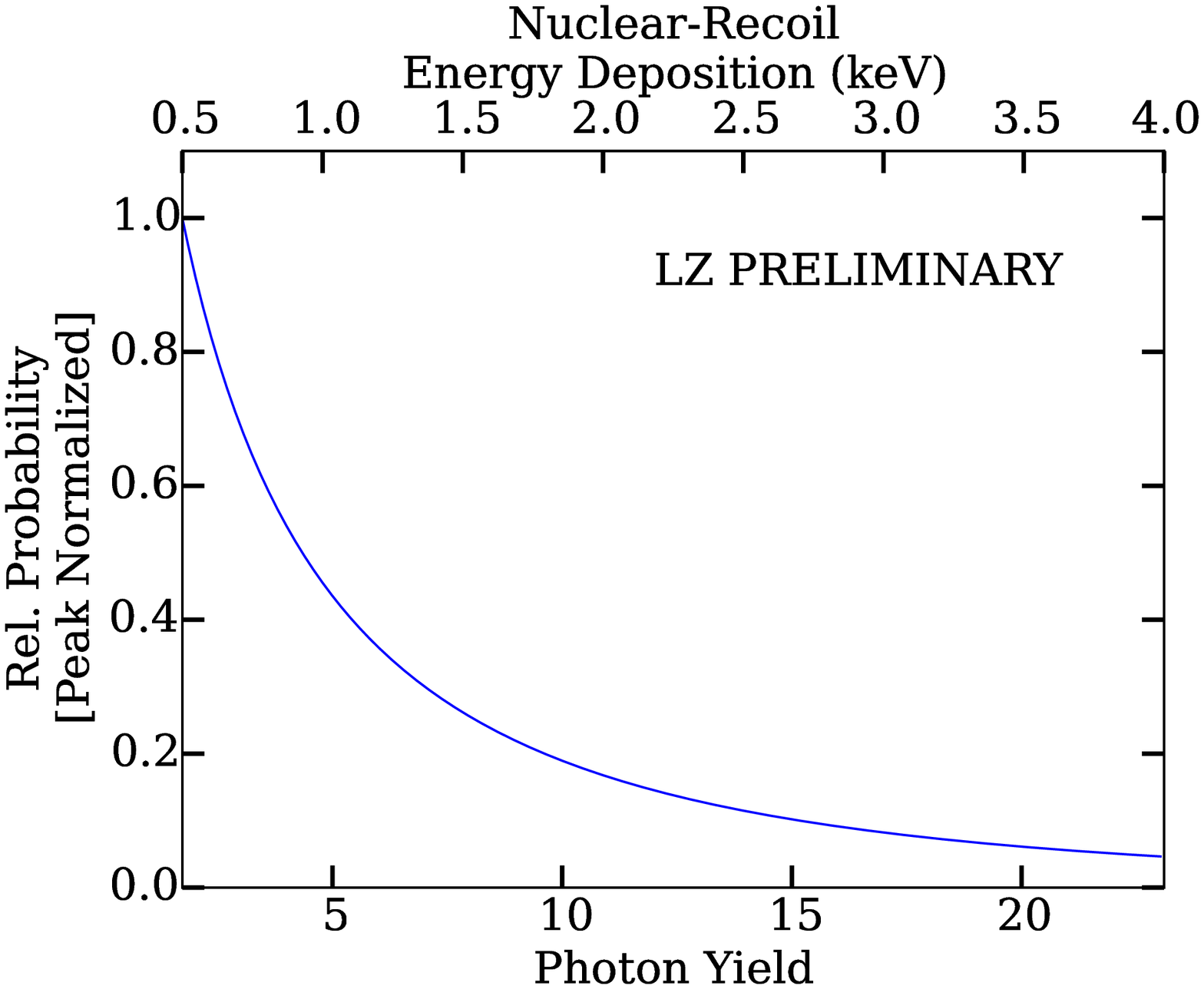}
\caption{The electron \emph{(Left)} and photon \emph{(Right)} yield, predicted by NEST, for an incident neutrino spectrum with an average energy of 10~MeV and a spectral profile from the accretion phase.}
\label{fig:electron_photon_yield}
\end{figure}

LZ is a dual-phase liquid xenon time projection chamber (TPC) to be operated deep underground (4,300 meters water equivalent) at the Sanford Underground Research Facility~\cite{LZTDR,SURF}. The sensitive volume of the detector has a diameter of 1.456~m and height of 1.456~m and will contain 7 tonnes of liquid xenon. Each end of the TPC is instrumented with an array of Hamamatsu R11410-20 PMTs; 253 at the top in the gas phase and 241 at the bottom in the liquid. Signals from the PMTs, after amplification and shaping, are fed into custom digitization and data sparsification boards where a real-time trigger decision is made.

A particle interacting with the liquid xenon results in both scintillation and ionization. Reference~\cite{LZTDR} estimates a baseline scintillation light collection efficiency of $\sim$9\% in the PMTs, producing prompt signal denoted S1. The ionization electrons drift under the applied electric field to the liquid-gas interface and, within the maximum drift time of $<$1~ms, are extracted into the gas phase with efficiency of $\sim$97.6\%. Each extracted electron will produce an electroluminescence signal, denoted S2, with an average of $\sim$60 detected photons. If a signal greater than 0.5~keV (corresponding to $\sim$3~electrons) is detected, a trigger is generated and the PMT waveforms between 1.5 ms pre-trigger and 2.5-ms post-trigger are written to disk.

The O(10~MeV) neutrinos emitted from SN will deposit O(1~keV) nuclear recoils in liquid xenon~\cite{ref3, Drukier, Horowitz, KateScholbergNu, RafaelLZ}. Figure~\ref{fig:energy_dep} presents the neutrino energy deposition spectrum and the number of expected interactions for a 27~M$_\odot$ SN at 10~kpc. The resulting distributions of the number of ionization electrons and photons were calculated based on the NEST model~\cite{NEST,LUXDD}. An example of these distributions for the accretion phase is presented in Fig.~\ref{fig:electron_photon_yield}. To achieve the lowest possible detection threshold, an S2-only analysis is pursued. At the low energies favored for CE$\nu$NS interactions, the low light yield will hinder particle particle identification using the charge-to-light ratio.

The LZ DAQ has a double buffer system, implemented on Xilinx Kintex 7 FPGAs, to capture data . While data from one buffer is being written to disk, the other buffer is capturing new interaction data and awaiting a trigger. When a trigger is generated, only the data from interactions that occurred in the pre- and post-trigger windows are written to disk. A buffer can hold a maximum of $\sim$45~interactions. To write a single interaction to disk it will take $\sim$1.3~ms and thus to offload the maximum number of 45~interactions it will take $\sim$59~ms. Only data from one buffer can be written to disk at a time. If both buffers are full and an additional trigger arrives, those interactions will be lost. In addition to reporting the time of the S2 as the trigger time, the DAQ will also report the deadtime when no interactions are captured.

\begin{figure}[htbp]
\centering
\includegraphics[width=.45\textwidth]{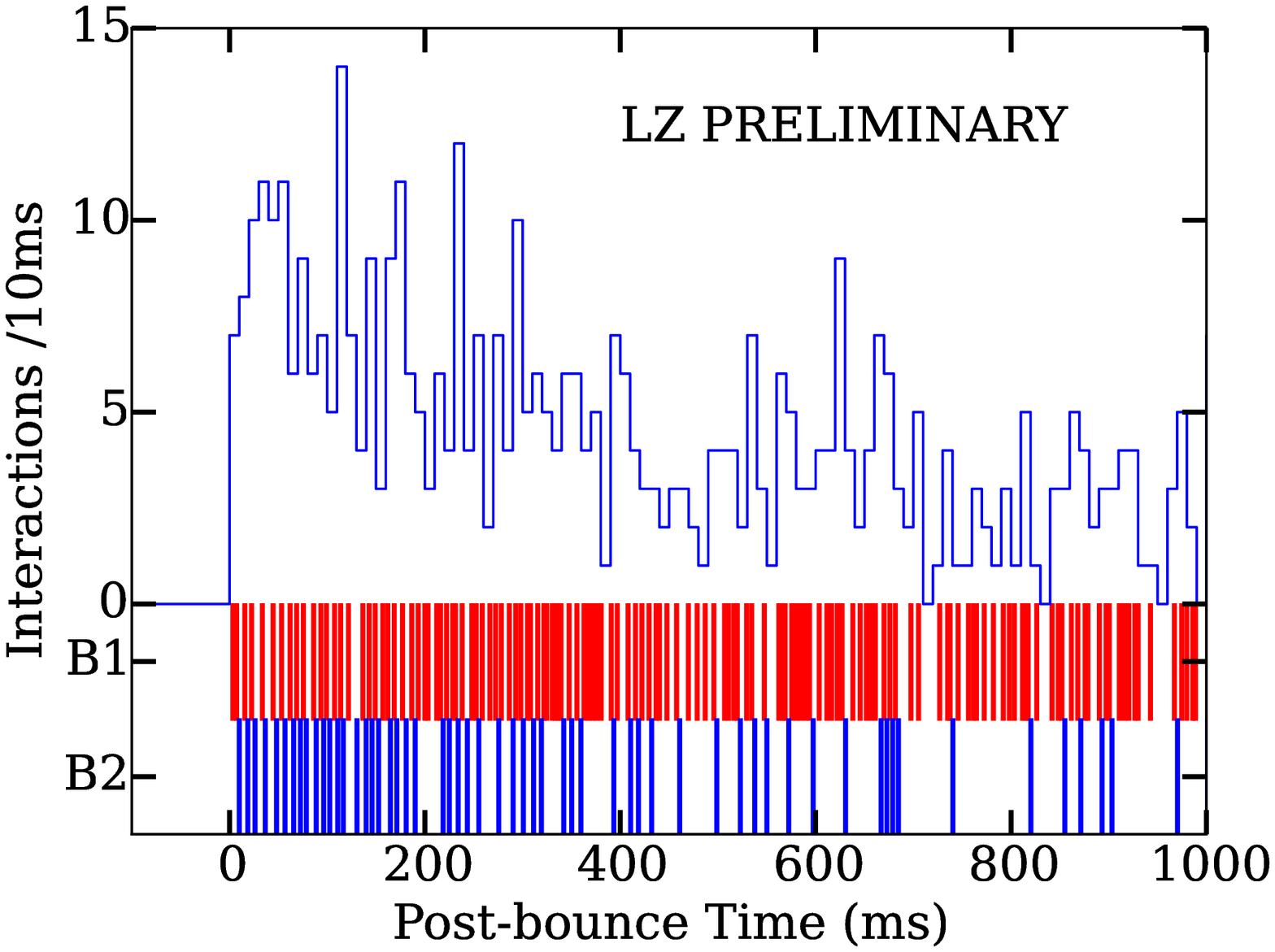}
\qquad
\includegraphics[width=.45\textwidth]{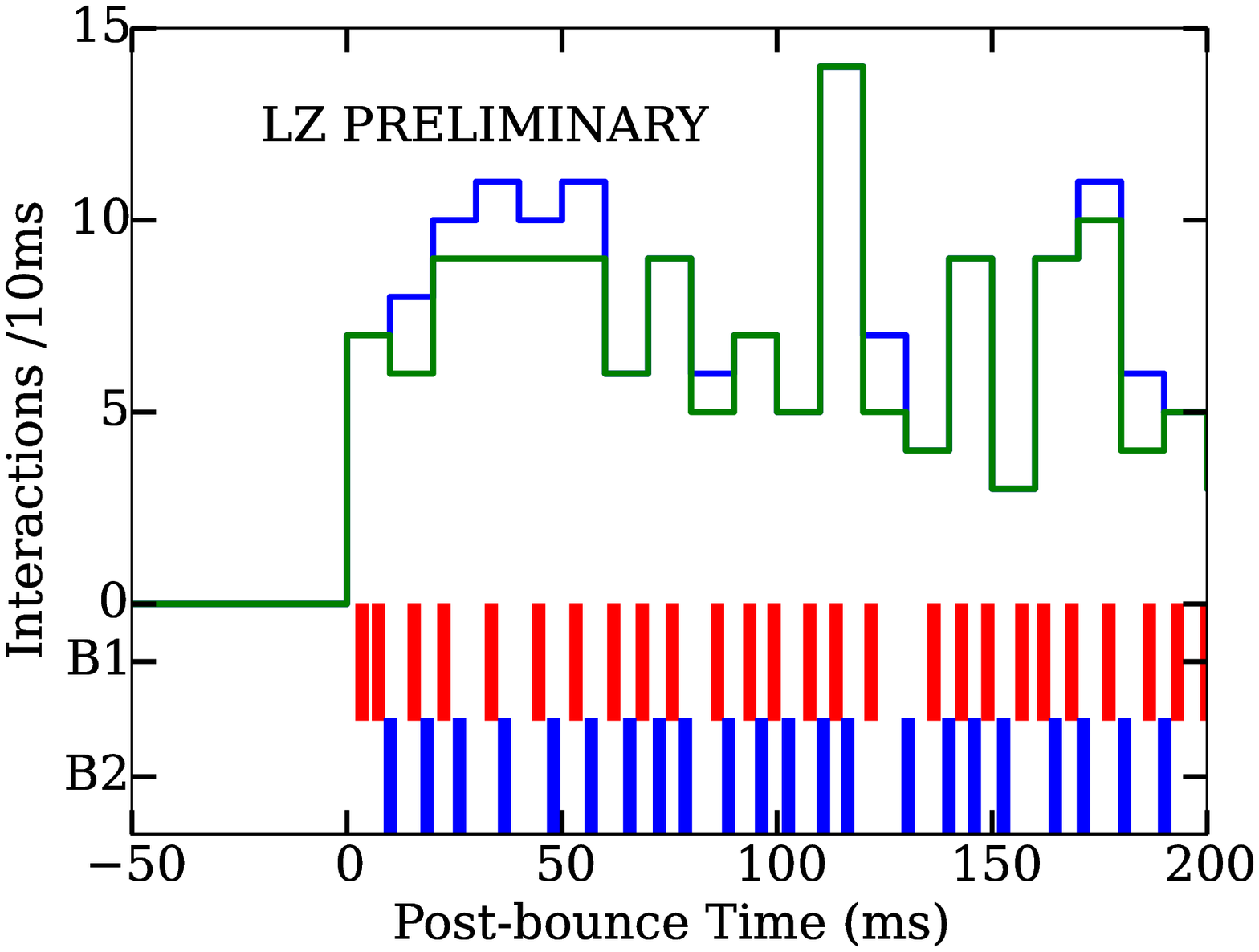}
\caption{\emph{(Left)} A time profile for a simulated SN event in LZ. Below the profile the bands indicated by `B1' and `B2' indicate the times when DAQ Buffer 1 and DAQ Buffer 2 were active. \emph{(Left)} For the 1~s of the event shown, there were 42 background interactions, 447 neutrino interactions and 470 out of the 489 total number of interactions were recorded. \emph{(Right)} A zoom in on the first 200~ms of the time profile that indicates the total number of interaction \emph{(blue)} and those that were recorded \emph{(green)}. It is during this phase of the SN that most of the neutrino interactions are missed.}
\label{fig:daq}
\end{figure}

We simulate how the LZ DAQ performs in response to triggers that have a time profile that matches the SN neutrino flux, summed over all flavors each with their own spectral energy distribution. A simulation is 11 seconds, 1 second before the SN bounce and 10 seconds post-bounce. In addition to the SN neutrino flux, we simulate a 40 Hz background interaction rate. Using the expected neutrino interaction distribution and background interaction rate we construct a composite probability distribution. We carry out Monte Carlo simulations where we vary the distance to the 27~M$_\odot$ SN. The distance to the SN is a proxy for the number of neutrino interactions detected. At each distance we allow the number of detected neutrino interactions to vary according to Poisson statistics. The times for these interactions is drawn from the probability distribution. These interaction times are then processed using the DAQ algorithm to determine which interactions are written to disk. An example of this is shown in Fig.~\ref{fig:daq}. 

\section{Trigger Time Analysis}
\label{sec:analysis}

\begin{figure}[htbp]
\centering 
\includegraphics[width=.45\textwidth]{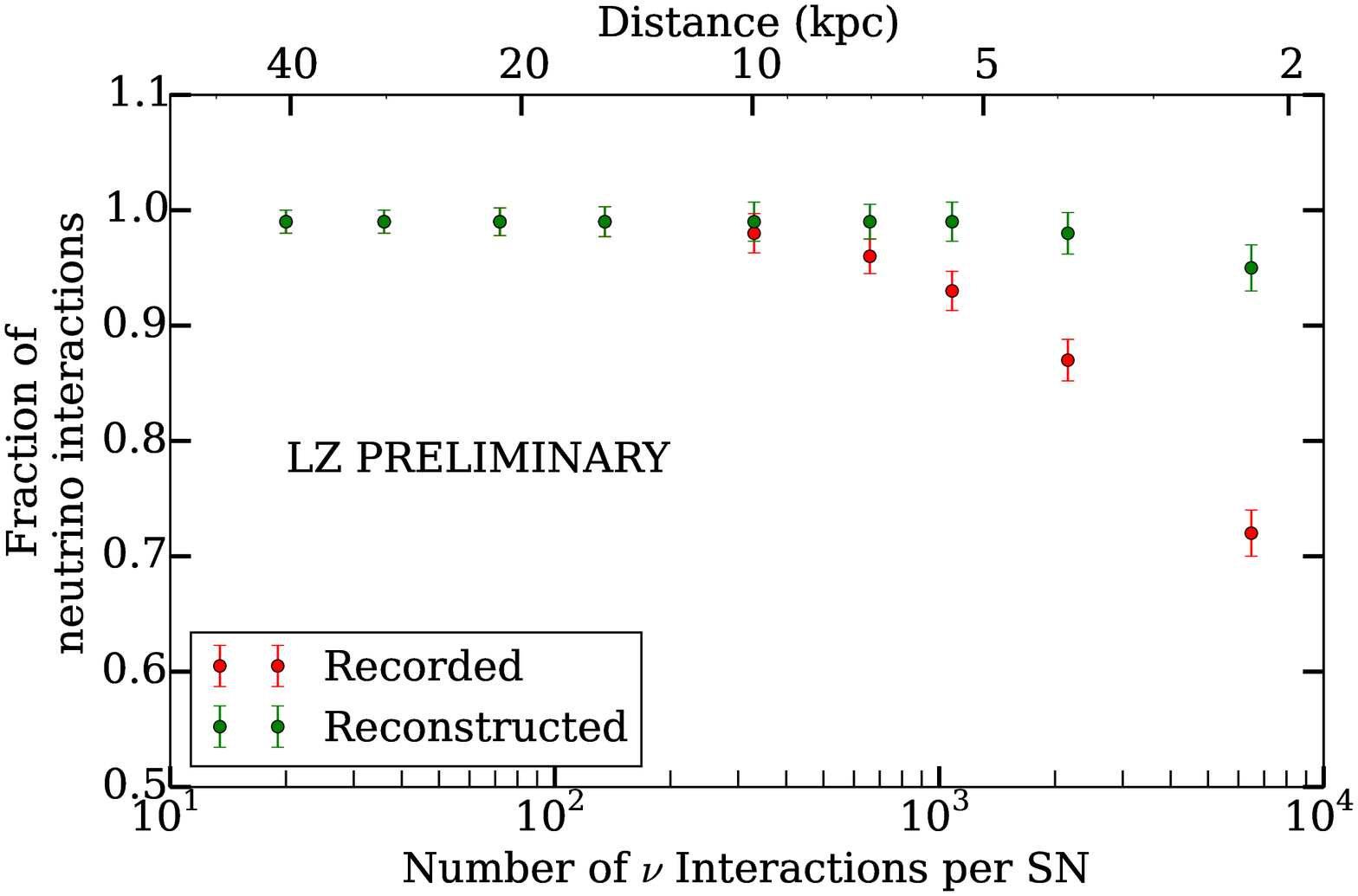}
\qquad
\includegraphics[width=.45\textwidth]{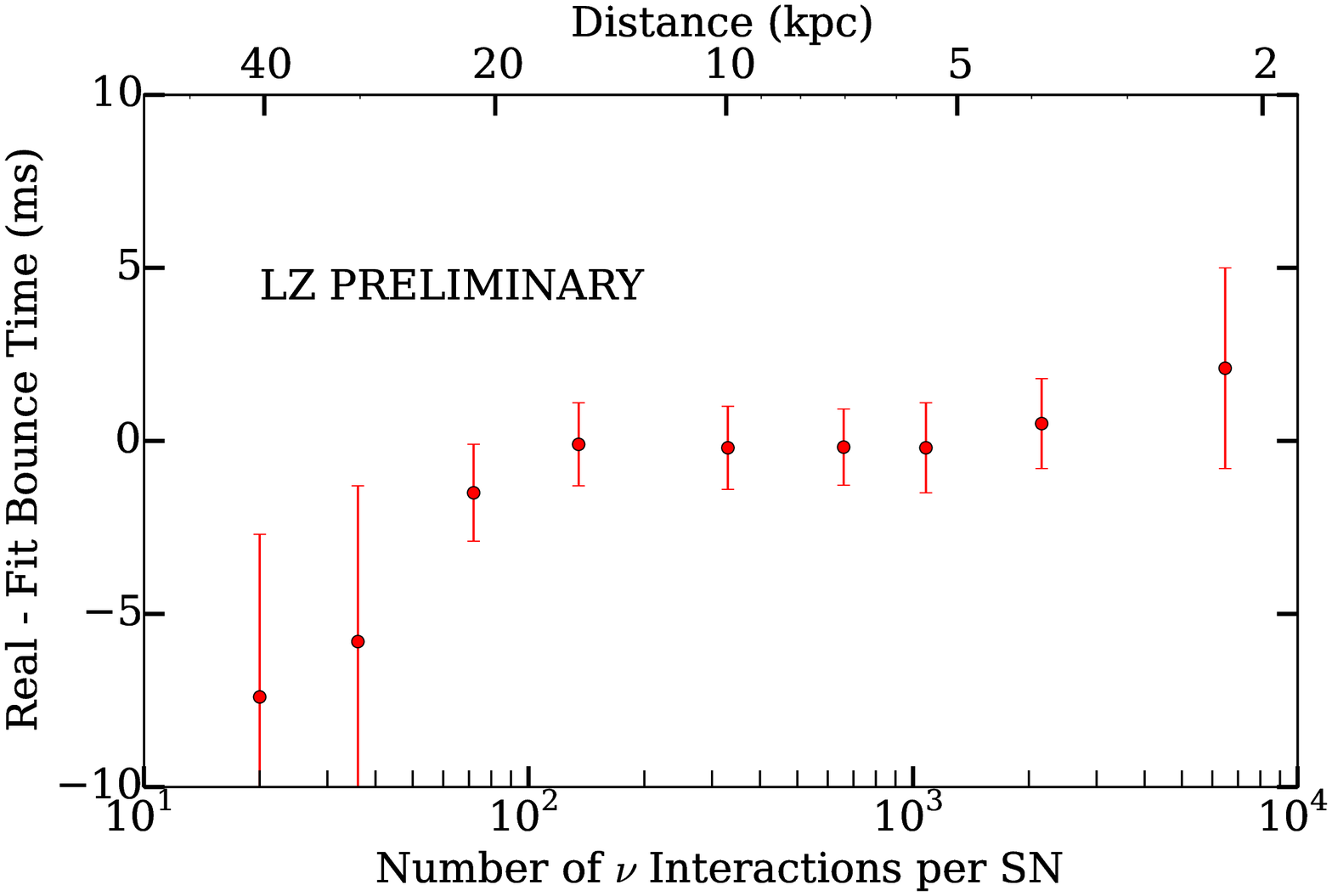}
\caption{\emph{(Left)} Fraction of recorded \emph{(red)} and reconstructed \emph{(green)} neutrino interactions. This fraction is presented versus number of neutrino interactions and distance from a 27 M$_\odot$ SN. \emph{(Right)} Reconstructed bounce time of the SN. The error bars in both plots indicate the statistical error associated with the simulated population.}
\label{fig:recon}
\end{figure}

Using the recorded trigger time information and the DAQ deadtime for an SN event, the true number of interactions that occurred and the SN's bounce time can be reconstructed. Since most SN events will be low interaction rate events an unbinned analysis is preferred to prevent biasing results. To accomplish this, the one dimensional trigger time information is integrated to give a cumulative number of recorded interactions as a function of time. With the known neutrino flux model and a uniform distribution of background events a composite model is created. It is then corrected for DAQ deadtime by removing those periods from the composite model. This model is compared to the cumulative interaction time data and the least squares is computed. We allow the bounce time of the model to vary to minimize the least squares. Models that cover a range of number of neutrino and background interactions are generated and are also compared to the data. The model that best fits the data is used to report the true interaction rate and bounce time of the SN.

The number of neutrino interactions in LZ goes as the inverse of distance squared. As the SN gets closer, the fraction of recorded interactions decreases, as is shown in Fig.~\ref{fig:recon}. Using the trigger time fitting method the true number of neutrino events can be reconstructed. We are able to accurately reconstruct the number of neutrino interactions within $>$90\% of the true number. Even in cases of low interaction rate (20 or 50 neutrinos) the model prefers to include neutrinos rather than one composed only of background. Also presented in Fig.~\ref{fig:recon} is the capability of the fitting method to reconstruct the bounce time of the SN. Low interaction rate events bias the bounce time to later times as there often isn't an interaction detected right at the onset of neutronization, neither is a well defined neutronization peak visible. However, we are able to still reconstruct the bounce time to within 10~ms. For high interaction rate events the two buffers are saturated with data early on and are dead for extended periods of time. This accounts for the large fraction of interactions missed and why the reconstructed time is earlier than the bounce time.

\section{Conclusions}
\label{sec:conc}

We have presented an overview of SN neutrino detection in LZ. These neutrinos are emitted with O(10~MeV) kinetic energy and will deposit O(1~keV) in liquid xenon. Using the known SN neutrino profile and a 40 Hz background we construct a Monte Carlo event generator. Assuming a 0.5~keV or $\sim$3~electron detection threshold for a trigger we simulate the response of the LZ DAQ's double buffer system to these events. The double buffering system of the DAQ will allow a large fraction of the SN neutrino interactions to be recorded. For a 27~M$_\odot$ SN at 10~kpc we expect 357~$\pm$~19 neutrino interactions. A SN 1987A type supernova at 50 kpc would produce ~20  interactions in LZ. We have developed a fitting routine that allows us to reconstruct the number of missed neutrino interactions and the bounce time of the SN. Even in low interaction rate SN events the fitting routine correctly picks out a model composed of an SN signal and background over a model of just the background. For high interaction rate SN events, we are able to reconstruct the number of neutrino interactions to $>$90\% of their true number. In all cases we fit the SN's bounce time to within 10~ms. 

\begin{acknowledgments}
The LZ experiment is partially supported by the U.S. Department of Energy (DOE) under award numbers DE-SC0012704, DE-SC0010010, DE-AC02-05CH11231, DE-SC0012161, DESC0014223, DEFG02-13ER42020, DE-FG02-91ER40674, DE-NA0000979, DE-SC0011702, DESC0006572, DE-SC0012034, DE-SC0006605, and DE-FG02-10ER46709; by the U.S. National Science Foundation (NSF) under award numbers NSF PHY-110447, NSF PHY-1506068, NSF PHY-1312561, and NSF PHY-1406943; by the U.K. Science \& Technology Facilities Council under award numbers ST/K006428/1, ST/M003655/1, ST/M003981/1, ST/M003744/1, ST/M003639/1, ST/M003604/1, and ST/M003469/1; and by the Portuguese Foundation for Science and Technology (FCT) under award numbers CERN/FP/123610/2011 and PTDC/FISNUC/1525/2014.
\end{acknowledgments}




\begin{thebibliography}{99}

\bibitem{ref1}
H.-T. Janka, T. Melson and A. Summa, \emph{Physics of Core-Collapse Supernovae in Three Dimensions: A Sneak Preview}, Annu. Rev. Nucl. Part. Sci. 66 (2016) 341-375.

\bibitem{ref2}
H.-T. Janka, \emph{Explosion Mechanisms of Core-Collapse Supernovae}, Annu. Rev. Nucl. Part. Sci. 62 (2012) 407-451.

\bibitem{ref3}
A. Alessandro Mirizzi, I. Tamborra, H.-T. Janka, N. Saviano, K. Scholberg, R. Bollig et al., \emph{Supernova neutrinos: Production, oscillations and detection}, Rev. Nuovo Cimento 39 (2016) 1-112.

\bibitem{ref4}
S. Chakraborty, R. Hansen, I. Izaguirre and G. Raffelt, \emph{Collective neutrino flavor conversion: Recent developments}, Nucl. Phys. B 908 (2016) 366-381.

\bibitem{ref5}
A. Mezzacappa, \emph{Ascertaining the core collapse supernova mechanism: The state of the art and the road ahead}, Annu. Rev. Nucl. Part. Sci. 55 (2005) 467–515.

\bibitem{ref6}
K. Scholberg, \emph{Future underground large detectors: prospects and physics case}, J. Phys. Conf. Ser. 375 (2012) 042048.

\bibitem{Drukier}
A. Drukier and L. Stodolsky, \emph{Principles and applications of a neutral-current detector for neutrino physics and astronomy}, Phys. Rev. D 30 (Dec, 1984) 2295–2309.

\bibitem{Beacom}
J. F. Beacom, W. M. Farr and P. Vogel, \emph{Detection of supernova neutrinos by neutrino-proton elastic
scattering}, Phys. Rev. D 66 (Aug, 2002) 033001.

\bibitem{Horowitz}
C. J. Horowitz, K. J. Coakley and D. N. McKinsey, \emph{Supernova observation via neutrino-nucleus elastic scattering in the CLEAN detector}, Phys. Rev. D 68 (Jul, 2003) 023005.

\bibitem{Coherent}
D. Akimov, J. B. Albert, P. An, C. Awe, P. S. Barbeau, B. Becker et al., \emph{Observation of coherent elastic neutrino-nucleus scattering}, Science (2017).

\bibitem{SnNuFutureChakraborty}
S. Chakraborty, P. Bhattacharjee and K. Kar, \emph{Observing supernova neutrino light curve in future dark matter detectors}, Phys. Rev. D 89 (Jan, 2014) 013011.

\bibitem{TimeRecon}
G. Pagliaroli, F. Vissani, E. Coccia and W. Fulgione, \emph{Neutrinos from Supernovae as a Trigger for Gravitational Wave Search}, Phys. Rev. Lett. 103 (Jul, 2009) 031102.

\bibitem{SuNuSignal1}
A. Burrows, K. Klein and R. Gandhi,\emph{The future of supernova neutrino detection}, Phys. Rev. D 45 (May, 1992) 3361-3385.

\bibitem{SuNuSignal2}
T. Totani, K. Sato, H. E. Dalhed and J. R. Wilson, \emph{Future Detection of Supernova Neutrino Burst and Explosion Mechanism}, Astrophys. J. 496 (1998) 216.

\bibitem{SuNuSignal3}
T. Fischer, S. Whitehouse, A. Mezzacappa, F.-K. Thielman and M. Liebendorfer, \emph{The neutrino signal from protoneutron star accretion and black hole formation}, Astron. Astrophys. 299 (2009) 1-15.

\bibitem{SuNuSignal4}
J. Gava, J. Kneller, C. Volpe and G. C. McLaughlin, \emph{Dynamical Collective Calculation of Supernova Neutrino Signals}, Phys. Rev. Lett. 103 (Aug, 2009) 071101.

\bibitem{KateScholbergNu}
K. Scholberg, \emph{Supernova neutrino detection}, Annu. Rev. Nucl. Part. Sci. 62 (2012) 81–103.

\bibitem{NuSpectFit}
I. Tamborra, B. Muller, L. Hudepohl, H.-T. Janka and G. Raffelt, \emph{High-resolution supernova neutrino spectra represented by a simple fit}, Phys. Rev. D 86 (Dec, 2012) 125031.

\bibitem{LS220EOS}
J. M. Lattimer and F. D. Swesty,\emph{A generalized equation of state for hot, dense matter}, Nucl. Phys. A 535 (1991) 331-376.

\bibitem{LZTDR}
B. J. Mount et al., \emph{LUX-ZEPLIN (LZ) Technical Design Report},arxiv:1703.09144.

\bibitem{SURF}
J. Heise, \emph{The Sanford Underground Research Facility at Homestake}, J. Phys. Conf. Ser. 606 (2015) 012015.

\bibitem{RafaelLZ}
R. F. Lang, C. McCabe, S. Reichard, M. Selvi and I. Tamborra, \emph{Supernova neutrino physics with xenon dark matter detectors: A timely perspective}, Phys. Rev. D 94 (Nov, 2016) 103009.

\bibitem{NEST}
M. Szydagis, N. Barry, K. Kazkaz, J. Mock, D. Stolp, M. Sweany et al., \emph{NEST: a comprehensive model for scintillation yield in liquid xenon}, J. Instrum. 6 (2011) P10002.

\bibitem{LUXDD}
LUX collaboration, D. S. Akerib et al., \emph{Low-energy (0.7-74 keV) nuclear recoil calibration of the LUX dark matter experiment using D-D neutron scattering kinematics}, arxiv: 1608.05381.


\end{thebibliography}
\end{document}